\documentclass[slac_one]{revtex4}
\usepackage{graphicx}
\usepackage{fancyhdr}
\newcommand{\MET}{\mbox{$\raisebox{.3ex}{$\not\!$}E_T$}}
\begin{document}

\title{High-Mass Resonances Decaying to Leptons and Photons at the Tevatron} 

%

\author{O. Stelzer-Chilton (on behalf of the CDF and D0 Collaborations)}
\affiliation{TRIUMF, Vancouver, BC, V6T 2A3, Canada}

\begin{abstract} 
The high-mass spectrum of lepton and photon pairs is sensitive to
a broad range of new physics. Examples are extra dimensions and
new gauge bosons such as the W' and Z'. Additionally, electron
compositeness would result in excited electrons that decay into
an electron and a photon. We report the latest results of searches
for high-mass dilepton, diphoton, and electron-photon resonances
by the CDF and D0 experiments at the Tevatron.
\end{abstract}

\maketitle

\thispagestyle{fancy}


\section{Introduction}
Searches for new physics that involve high momentum electrons, muons and photons in the final state provide a
clean environment to look for resonances at high mass. The standard model (SM) backgrounds are typically modeled with
Monte Carlo simulation with the exception of misidentified backgrounds from QCD processes that are usually derived directly from data.
Several experimental signatures will be discussed and in turn be interpreted in specific 
beyond the standard model theories.
\vspace{-0.5cm}
\section{Search for High Mass Di-Electron and Di-Photon Resonances}
The large difference between the Planck scale, $M_{Pl}$ $\approx$ $10^{16}$ TeV, and the weak scale presents a strong indication
that the standard model is incomplete. In the Randall-Sundrum (RS) scenario \cite{rs}, the space-time metric varies exponentially 
in a fourth spatial dimension. The wave function overlap with the SM brane is therefore suppressed, thus explaining 
the apparent weakness of gravity. This model predicts a tower of Kaluza-Klein excitations represented as massive graviton 
modes that couple with similar strength as the weak interaction. Their properties are quantified by two
parameters, the mass of the first massive excitation $M$ and the dimensionless coupling constant to standard model
fields, $k/\bar M_{Pl}$, where $\bar M_{Pl}=M_{Pl}/\sqrt(8\pi)$ is the reduced Planck scale.

\begin{figure}[h]
\begin{minipage}{19pc}
\includegraphics[height=12pc,width=20pc]{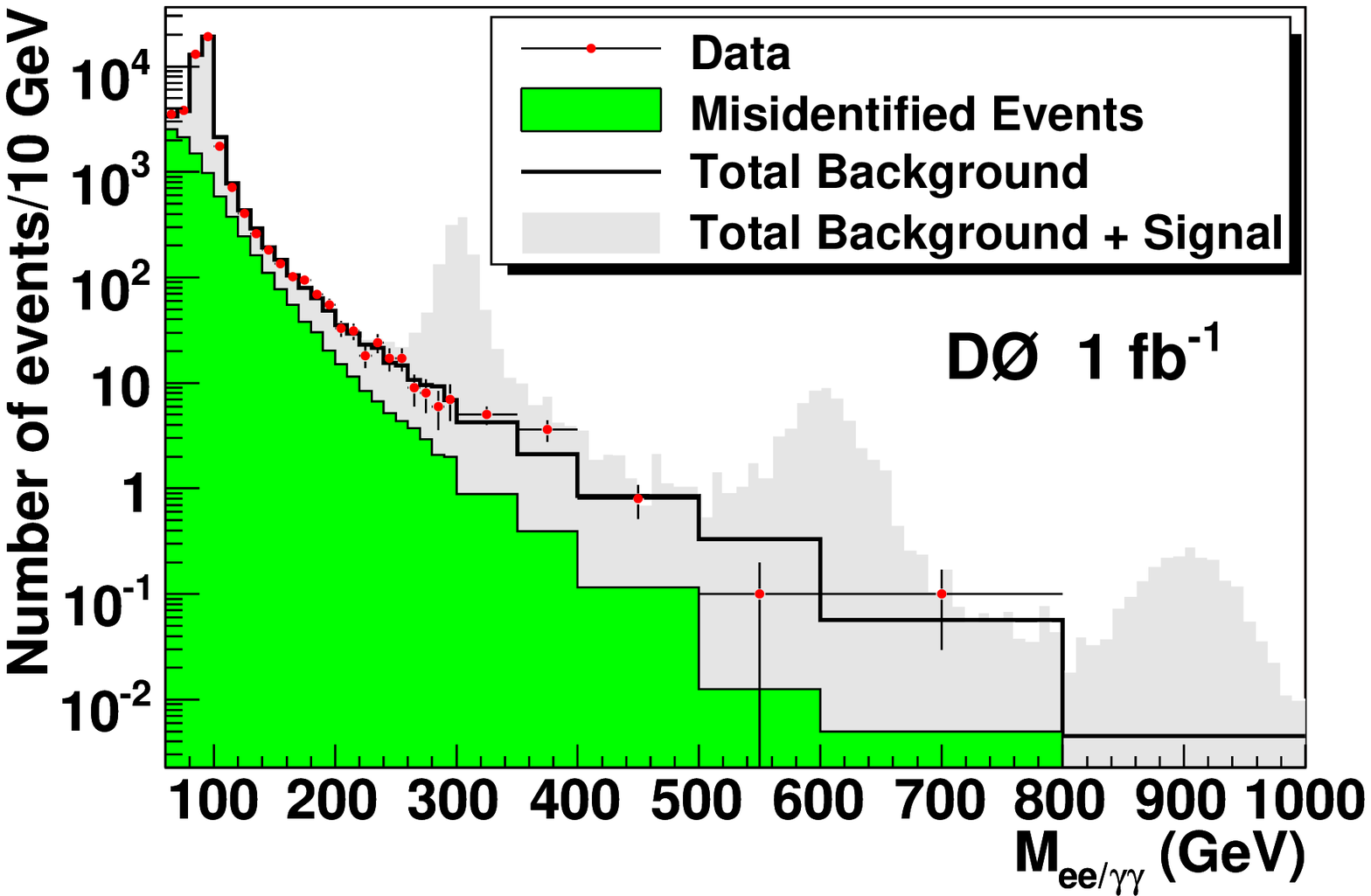}
\end{minipage}\hspace{1pc}%
\begin{minipage}{19pc}
\includegraphics[height=12pc,width=20pc]{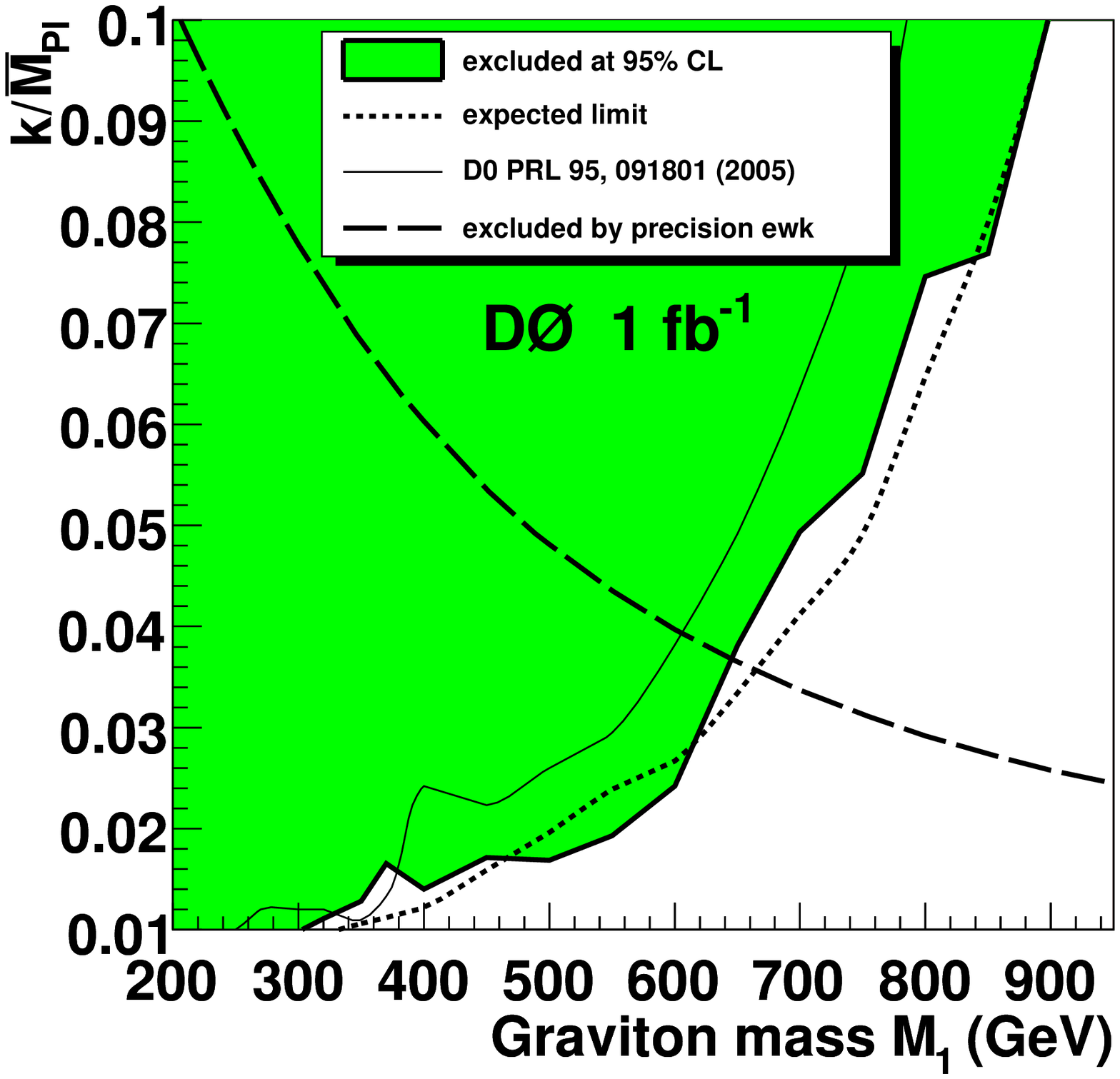}
\end{minipage}
\caption{Left: The invariant mass $M_{ee/\gamma\gamma}$ spectrum. Right: The 95\% C.L. excluded regions as a function of $M$ and $k/\bar M_{Pl}$. \label{d0_diele}}
\end{figure}

\begin{figure}[h]
\begin{minipage}{19pc}
\includegraphics[height=12pc,width=20pc]{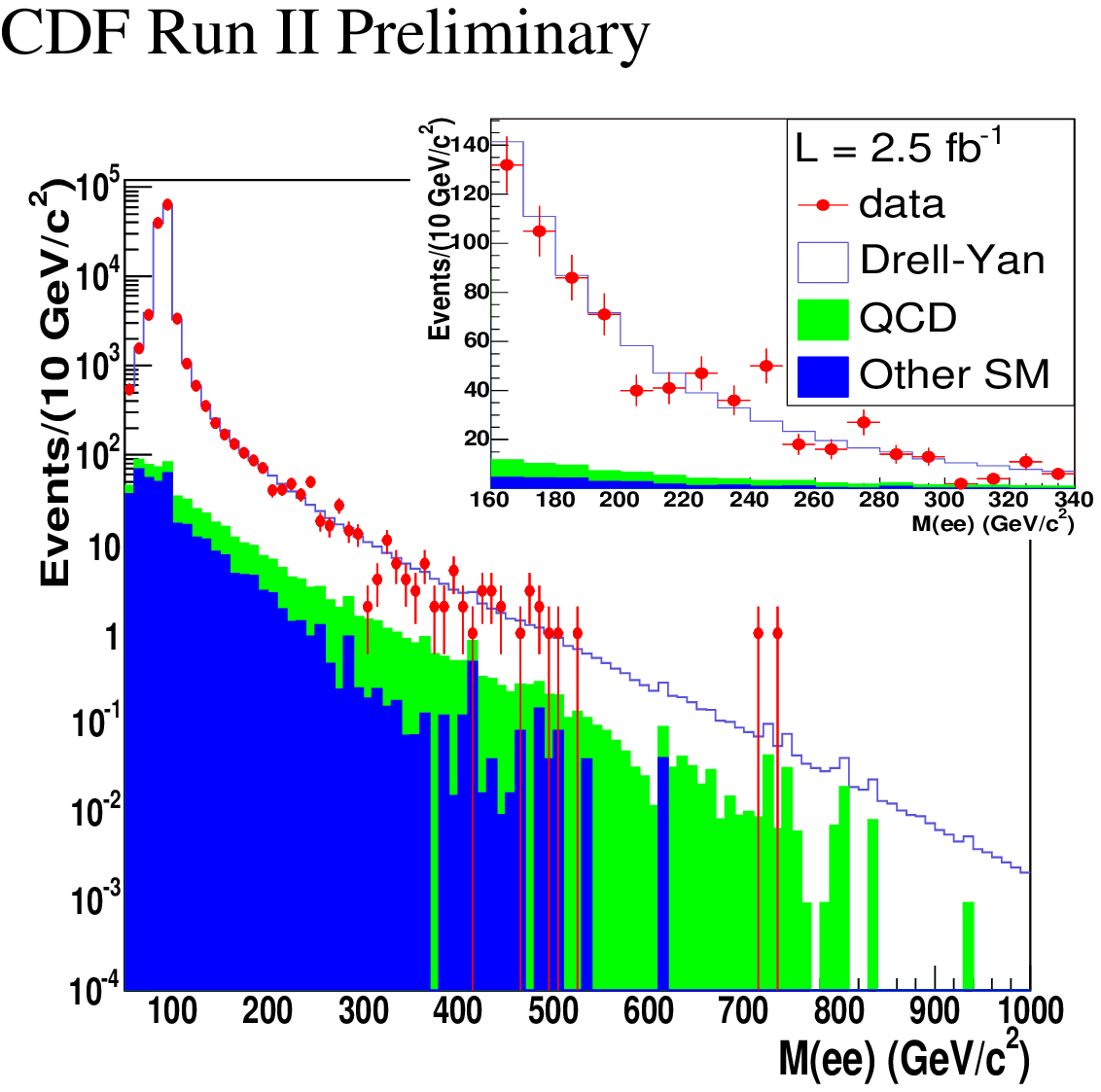}
\end{minipage}\hspace{1pc}%
\begin{minipage}{19pc}
\includegraphics[height=12pc,width=20pc]{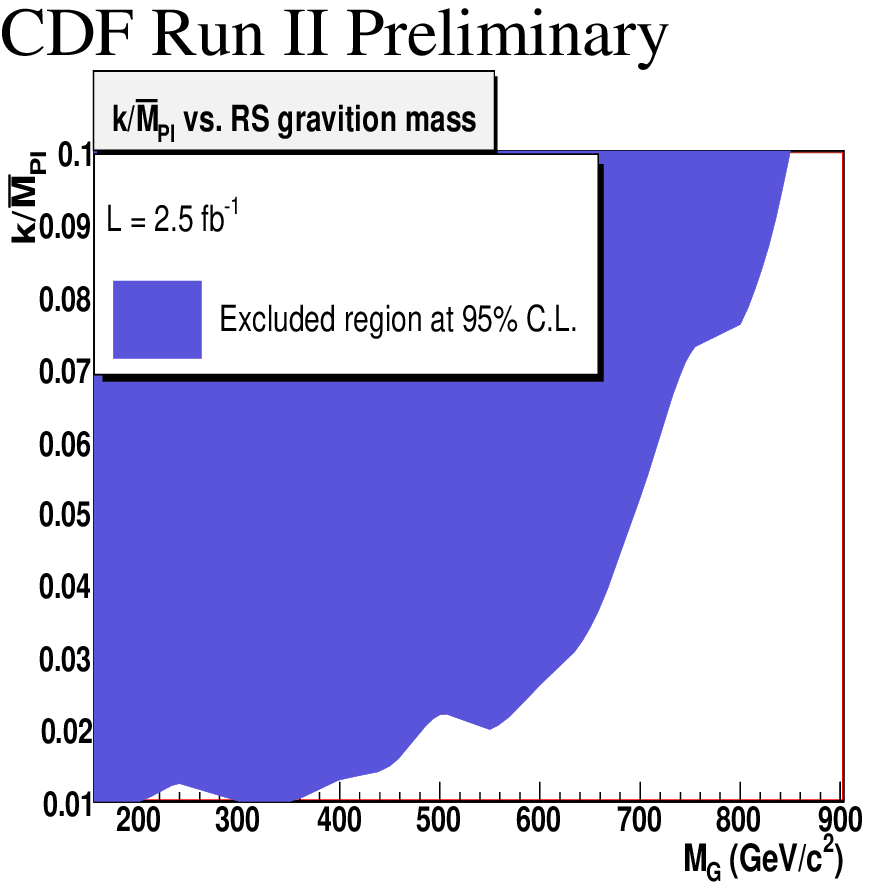}
\end{minipage}
\caption{Left: The invariant mass $M_{ee}$ spectrum. Right: The 95\% C.L. excluded regions as a function of $M$ and $k/\bar M_{Pl}$. \label{cdf_diele}}
\end{figure}

The CDF and D0 collaborations have searched for high mass resonances in the $ee$ \cite{cdf_rs_zprime} and $ee/\gamma\gamma$ \cite{d0_rs} final states,
using 2.5 and 1.0 fb$^{-1}$ of data, respectively. The CDF analysis requires two electrons in the central-central
($|\eta|<1.1$) or central-forward ($2.0>|\eta|>1.2$) region. The CDF electrons and the D0 electromagnetic (EM) clusters
must satisfy $E_T$$>$25 GeV. The left side of Figure~\ref{d0_diele} and ~\ref{cdf_diele} show the $M_{ee}$ and
$M_{ee/\gamma\gamma}$ spectra from CDF and D0. 
The D0 data are consistent with standard model predictions. The p-value of the largest excess in the CDF data 
at $228<M_{ee}<250$ GeV is 0.6\%. Without
significant excesses in both analyses, CDF and D0 set limits on the mass of RS gravitons as a function of coupling strength, as shown in
Figure ~\ref{d0_diele} and ~\ref{cdf_diele} on the right. For $k/\bar M_{Pl}=0.1$, masses below 850 GeV and 900 GeV are
excluded, for CDF and D0 respectively. Using the di-electron spectrum, CDF also sets limits for Z' bosons in various models,
as will be discussed in section 3.

\begin{figure}[!h]
\begin{minipage}{19pc}
\includegraphics[height=12pc,width=20pc]{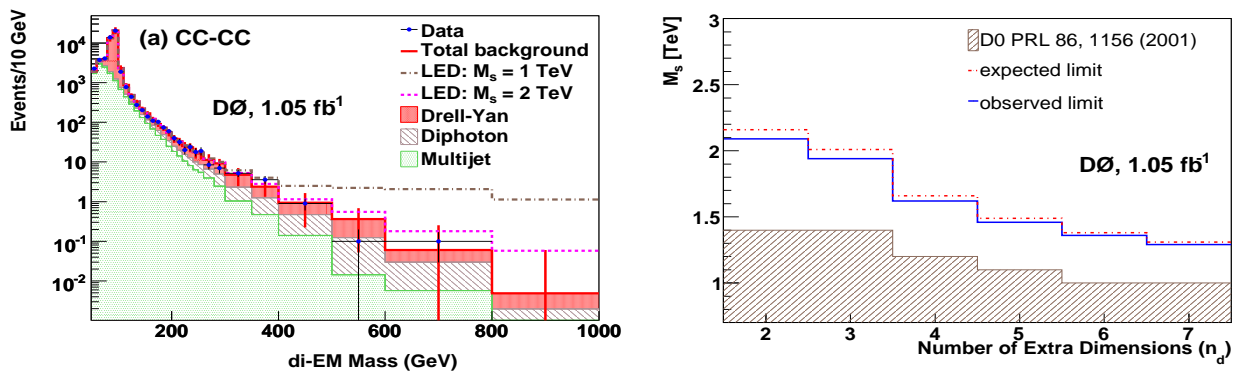}
\end{minipage}\hspace{1pc}%
\begin{minipage}{19pc}
\includegraphics[height=12pc,width=20pc]{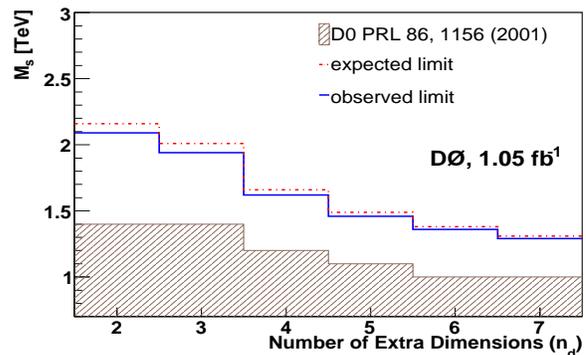}
\end{minipage}
\caption{Left: The invariant mass $M_{ee/\gamma\gamma}$ spectrum. Right: The 95\% C.L. limit on the effective Planck scale $M_S$ vs $n_d$. \label{d0_led}}
\end{figure}

An alternative way to circumvent the hierarchy problem is by extending the dimensionality of space, the approach
in the ADD large extra dimension model \cite{add}. This model posits that 
gravity propagates in $n_d$ additional 
compactified spatial dimensions. Gauss's Law gives the relation between the effective Planck scale $M_S$, the observed 
Planck scale, and the size of the extra dimensions R: $M_{Pl}^2 \sim R^{n_d}M_S^{n_d+2}$.
If R is large compared to the 
Planck length, $M_S$ can be as low as 1 TeV. Extra spatial dimensions will manifest themselves by the presence of a series of 
graviton states that will result in enhancement of the cross sections above the SM values, especially 
at high energies.
The D0 collaboration used 1.0 fb$^{-1}$ of data with a similar selection of a combined $ee/\gamma\gamma$ final state
as for the RS graviton search, but also including EM candidate showers in the forward calorimeters \cite{d0_add}. The data is shown
on the left side of Figure~\ref{d0_led} and is consistent with the background hypothesis.
The obtained limits as a function of $n_d$ are illustrated on the right side of Figure~\ref{d0_led}, which range from 1.29   
to 2.09 TeV at the 95\% C.L. for $n_d$ = 7 to $n_d$ = 2.
\vspace{-0.5cm}
\section{High Mass Di-Muon Spectrum} 
\begin{figure}[t]
\begin{minipage}{19pc}
\includegraphics[height=12pc,width=20pc]{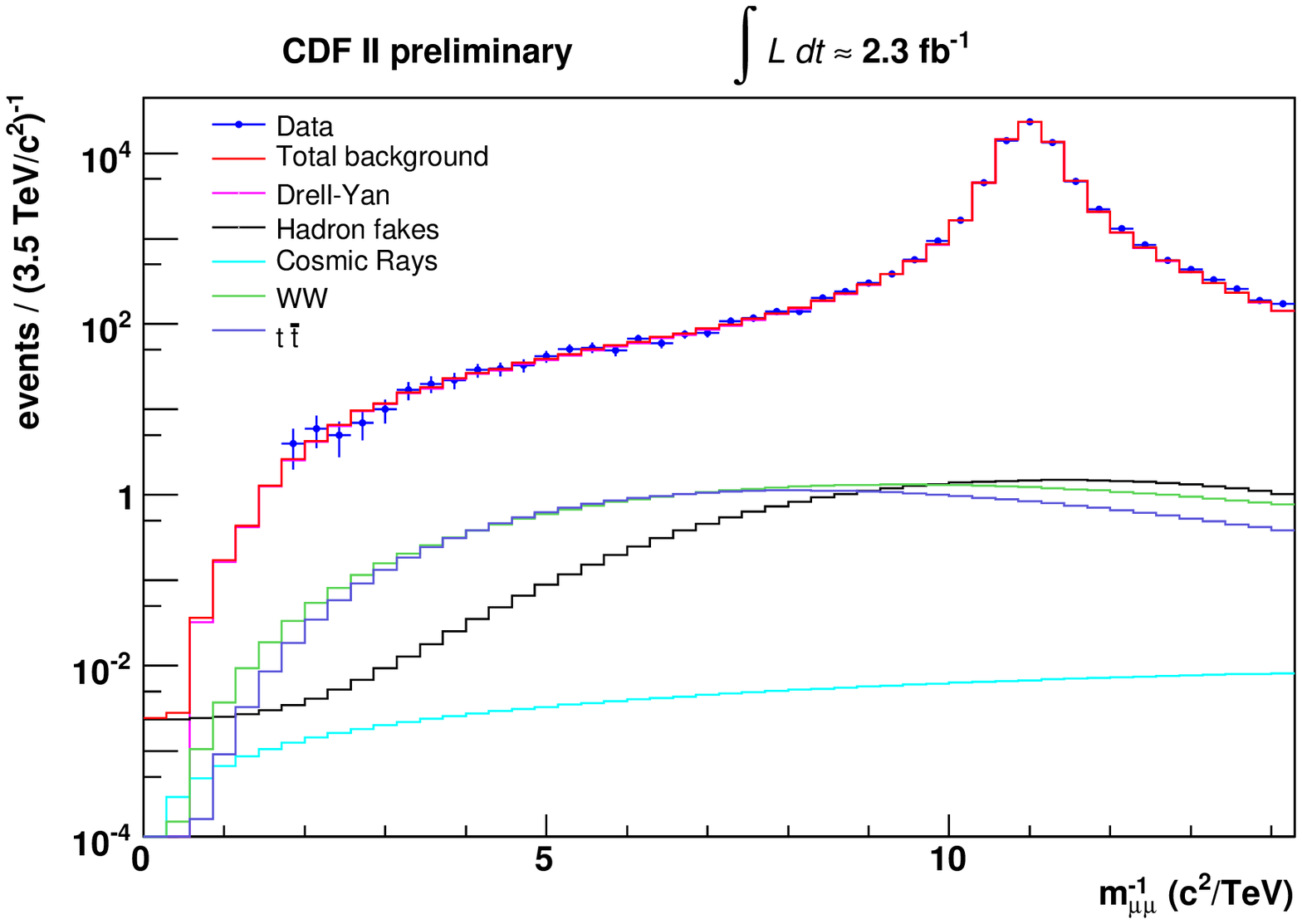}
\end{minipage}\hspace{1pc}%
\begin{minipage}{19pc}
\includegraphics[height=12pc,width=20pc]{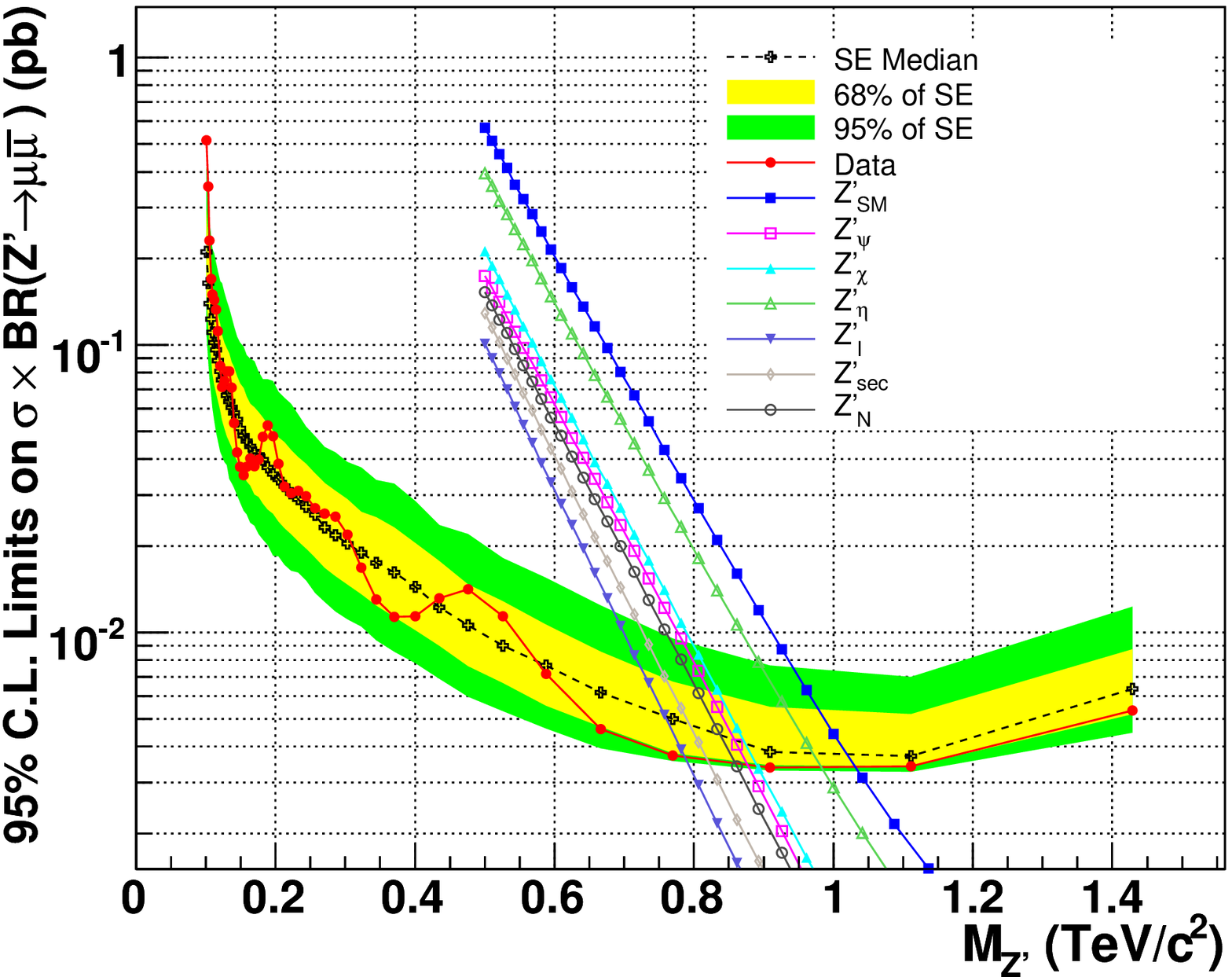}
\end{minipage}
\caption{Left: The inverse invariant mass $m_{\mu\mu}^{-1}$ spectrum. Right: The 95\% C.L. limit for various Z' couplings. \label{cdf_mu}}
\end{figure}

In many schemes of GUT symmetry-breaking \cite{zprime}, additional U(1) gauge groups survive to relatively low energies, 
leading to the prediction of additional neutral gauge vector bosons, generically referred to as Z' bosons. Such Z' bosons 
are expected to couple with electroweak strength to SM fermions, thus appearing as narrow, spin-1, resonances.
Many other models, such as the left-right model, and the little Higgs models, also predict heavy neutral gauge bosons.
A CDF search uses 2.3 fb$^{-1}$ of data and selects muons with a track $p_T$ $>$ 30 GeV. The analysis uses the inverse mass
spectrum as shown in Figure~\ref{cdf_mu}, with the benefit that the detector resolution is approximately constant over 
the range shown in the plot.   
The data is in good agreement with standard model expectations. 
Limits are set on Z' bosons  predicted by E6 models with different couplings or assuming the same couplings to SM fermions as the Z boson,
shown on the right side of Figure~\ref{cdf_mu}. Table I summarizes all the derived lower mass limts for the CDF di-muon search as well as the
CDF di-electron search described in section 1.
\begin{table}[h]
\begin{minipage}{19pc}
\begin{center}
\caption{Z' Mass Limits}
\begin{tabular}{l|c|c|c|c|c|c|c}
\textbf{Channel} & \textbf{$Z'_{I}$} & \textbf{$Z'_{sec}$} & \textbf{$Z'_{N}$} & \textbf{$Z'_{\psi}$} & \textbf{$Z'_{\chi}$} & \textbf{$Z'_{\eta}$} & \textbf{$Z'_{SM}$}
\\
\hline CDF $\mu\mu$ & 789 & 821 & 861 & 878 & 892 & 982 & 1030\\
\hline CDF $ee$ & 737 & 800 & 840 & 853 & 864 & 933 & 966\\
\end{tabular}
\label{l2ea4-t1}
\end{center}
\end{minipage}
\begin{minipage}{19pc}
\begin{center}
\caption{Graviton Mass Limits}
\begin{tabular}{l|c|c}
\textbf{Channel} & \textbf{$k/M_{Pl}=0.01$} & \textbf{$k/M_{Pl}=0.1$} 
\\
\hline CDF $\mu\mu$ & 293 & 921 \\
\hline CDF $ee$ & 358 & 850 \\
\hline D0 $ee/\gamma\gamma$ & 300 & 900 \\
\end{tabular}
\label{l2ea4-t1}
\end{center}
\end{minipage}
\end{table}
CDF also extracts limits for spin-2 RS gravitons using the di-muon final state. Table II summarizes the derived mass limits
from the CDF $\mu\mu$ final states and the from the CDF $ee$ and D0 $ee/\gamma\gamma$ final states discussed in section 1.
\vspace{-0.5cm}
\section{High Transverse Mass Electron-Neutrino Spectrum}
\begin{figure}[h]
\begin{minipage}{19pc}
\includegraphics[height=12pc,width=20pc]{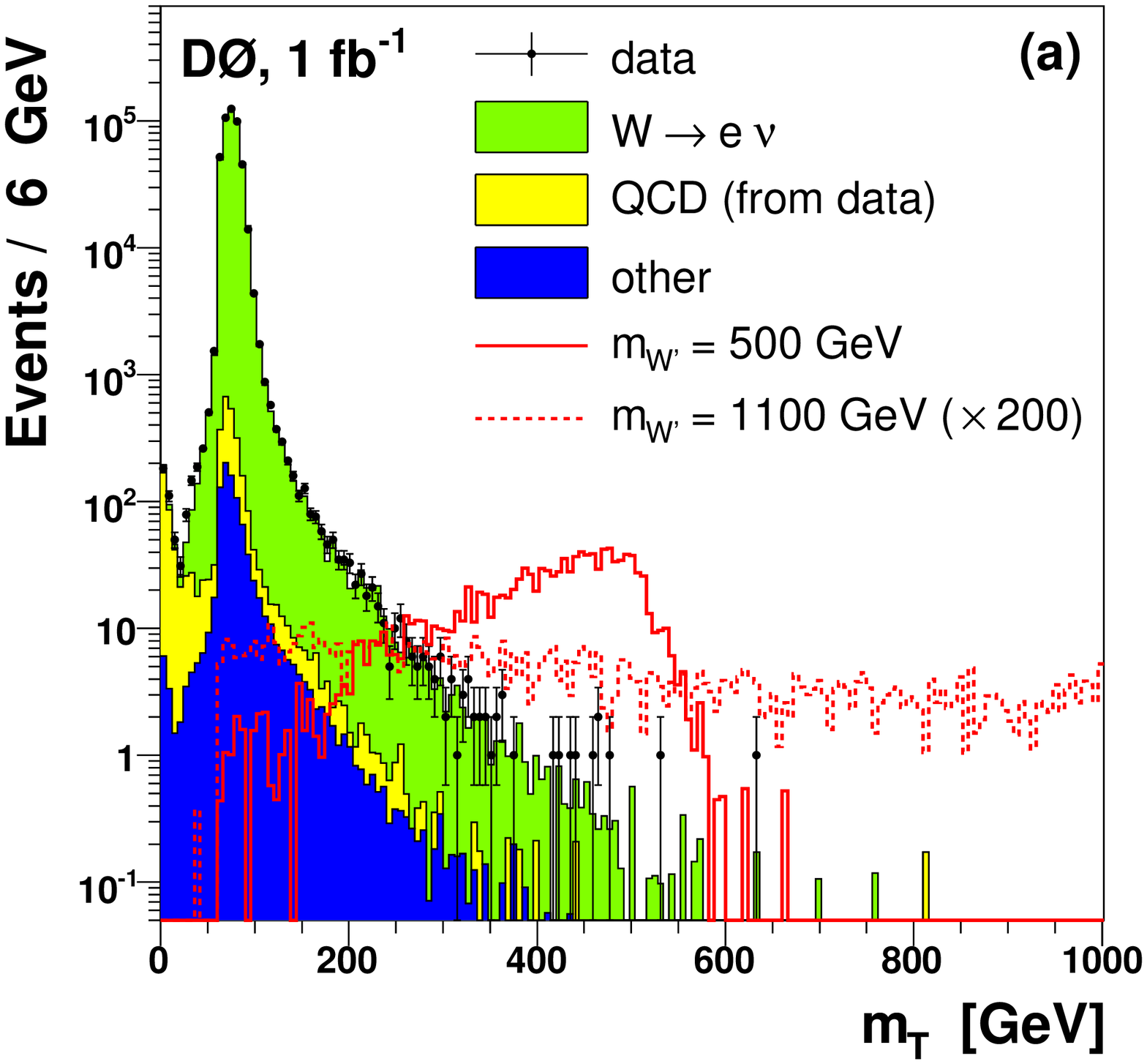}
\end{minipage}\hspace{1pc}%
\begin{minipage}{19pc}
\includegraphics[height=11pc,width=20pc]{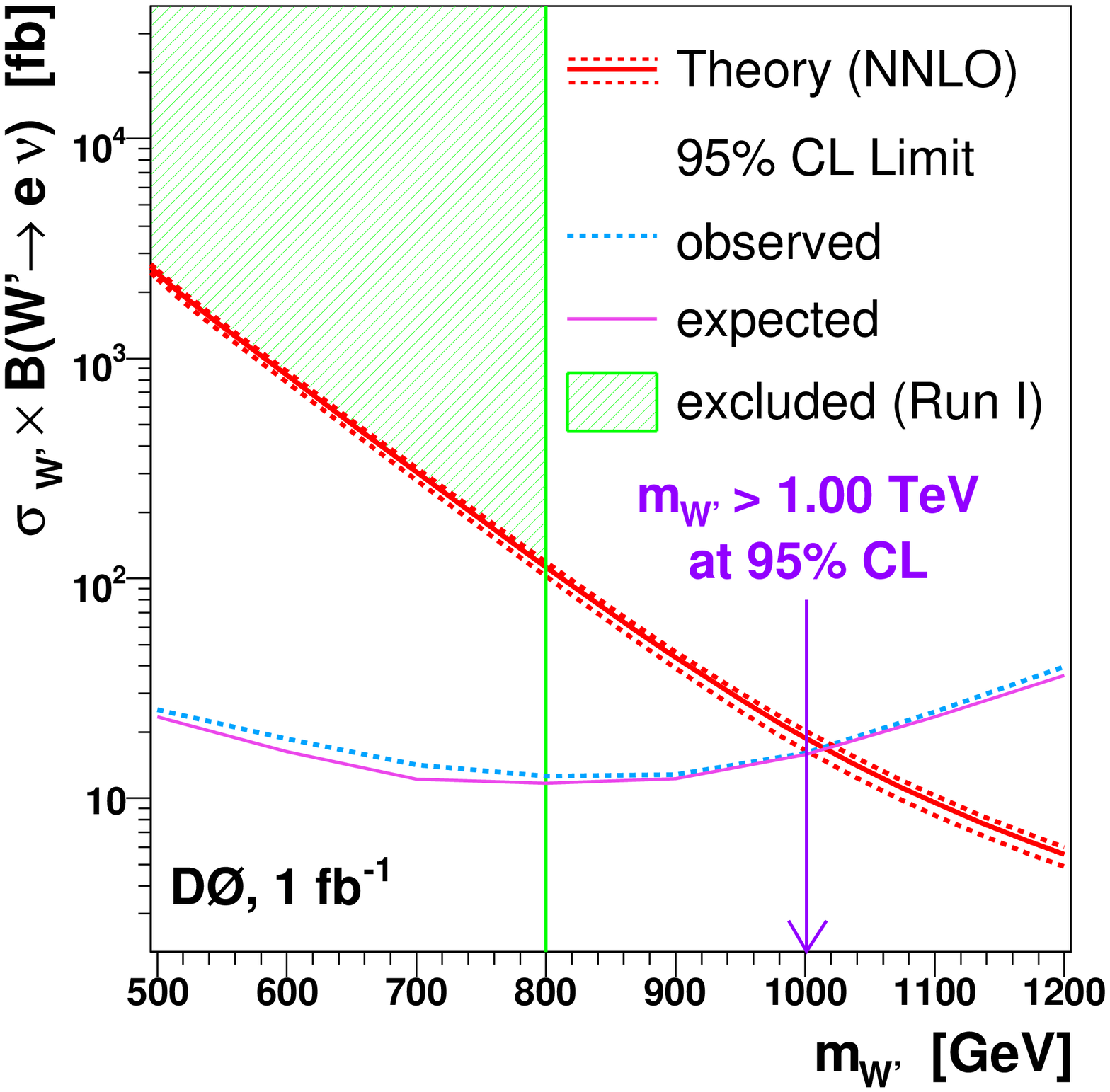}
\end{minipage}
\caption{Left: The transverse mass $m_T$ spectrum. Right: The 95\% C.L. limit as a function of W' mass. \label{d0_wprime}}
\end{figure}

Additional charged gauge bosons, W' bosons, have been introduced by several new physics models, such as left-right
symmetric and E6 models. The D0 collaboration has searched for a W' decaying to an electron and a neutrino using
1.0 fb$^{-1}$ of data \cite{d0_wprime}. Events are required to have a central electron with $E_T$$>$30 GeV and $\MET$$>$30 GeV. There
is no excess in the high transverse mass spectrum, shown on the left of Figure~\ref{d0_wprime}. 
Using the Altarelli reference model \cite{altarelli}, where SM couplings are assumed,
W' with masses below 1 TeV are excluded at 95\% CL, shown on the right side of Figure~\ref{d0_wprime}. 

\section{High Mass Electron-Photon Spectrum} 
\begin{figure}[h]
\begin{minipage}{19pc}
\includegraphics[height=12pc,width=20pc]{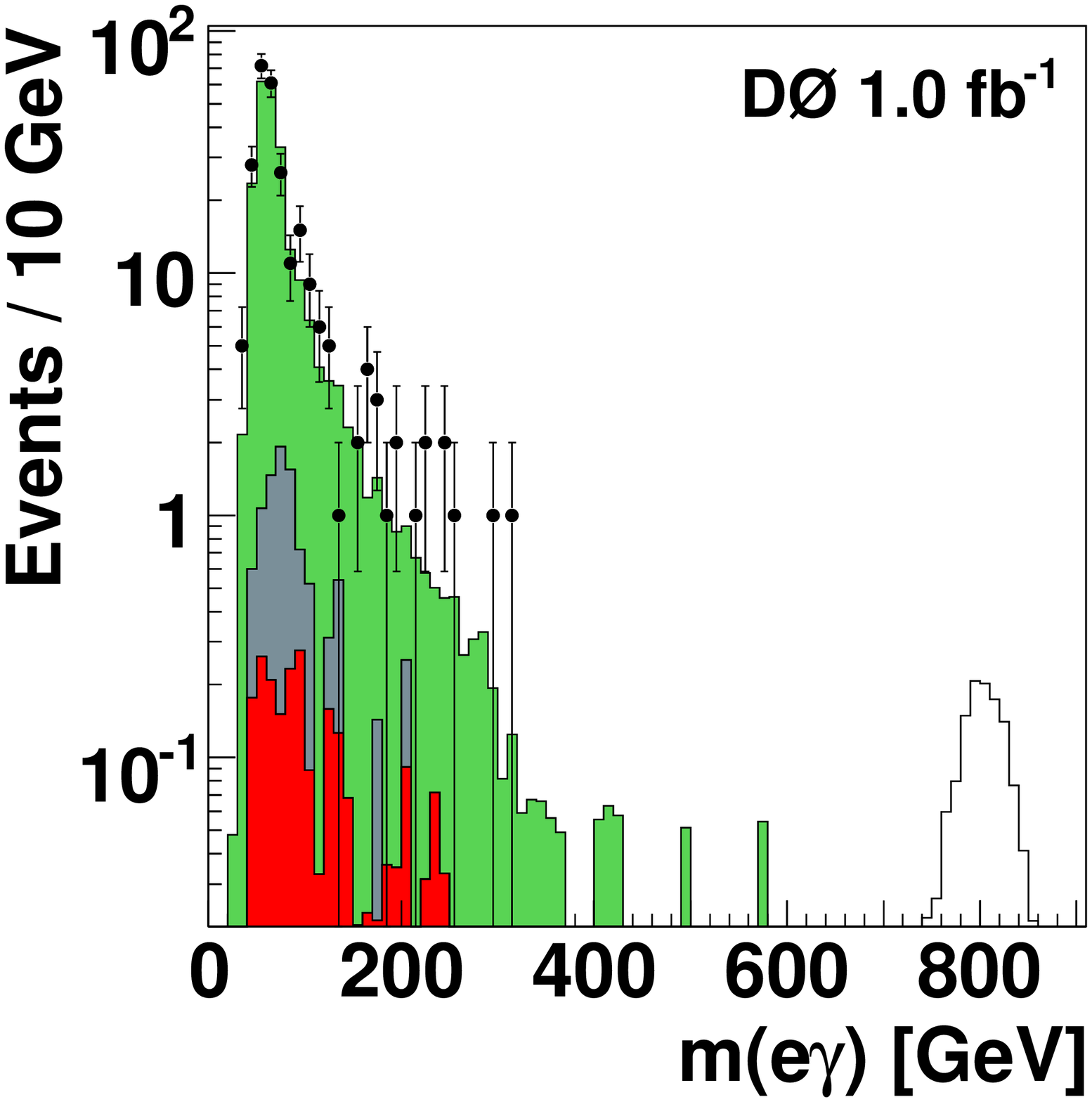}
\end{minipage}\hspace{1pc}%
\begin{minipage}{19pc}
\includegraphics[height=12pc,width=20pc]{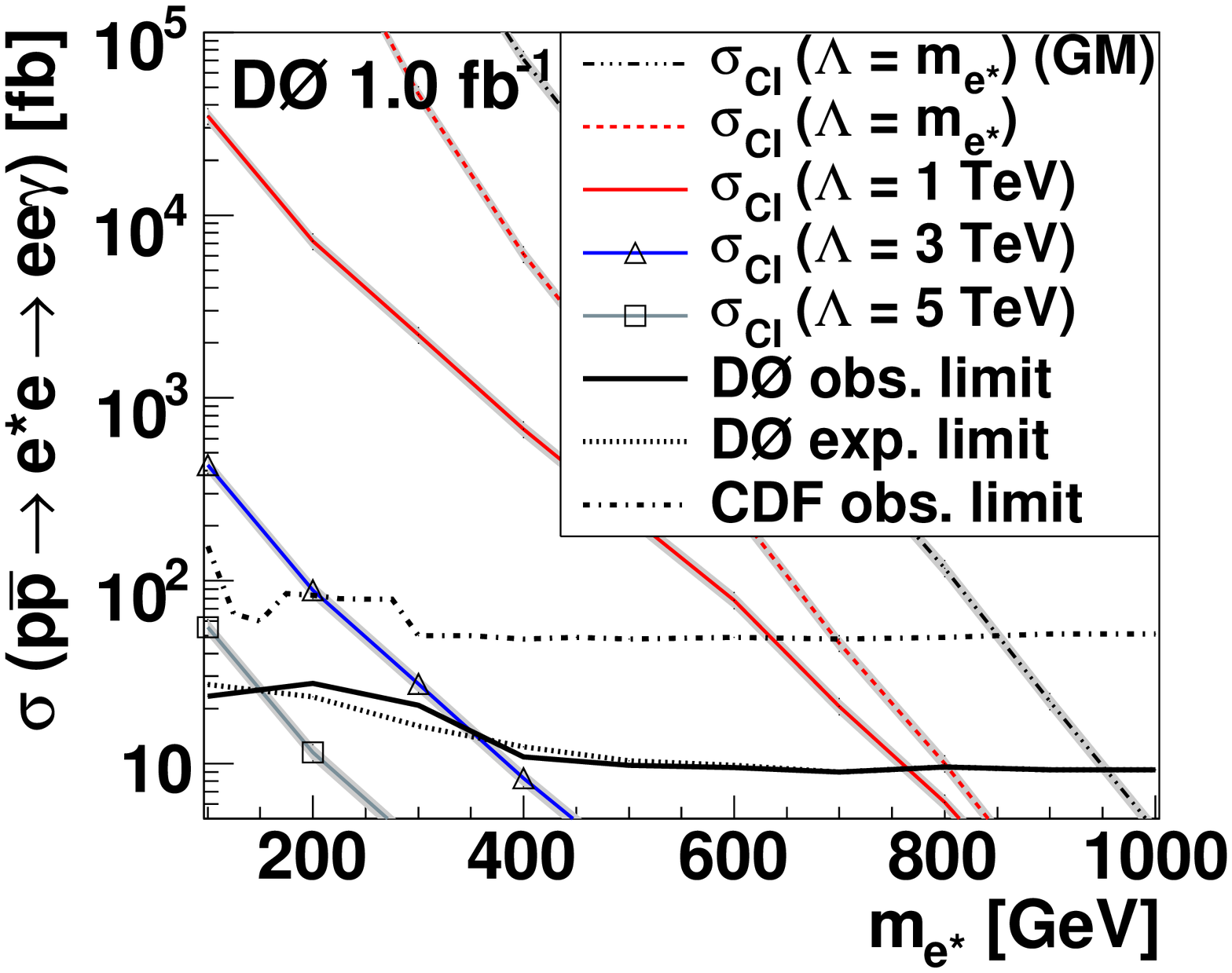}
\end{minipage}
\caption{Left: The invariant mass $m_{e\gamma}$ spectrum. Right: The 95\% C.L. limit as a function of contact interaction scale $\Lambda$. \label{d0_e*}}
\end{figure}

The observed fermion multiplicity motivates a description in terms of underlying substructure, in which all quarks and 
leptons consist of fewer elementary particles bound by a new strong interaction. In this compositeness model, 
quark-antiquark annihilations may result in the production of excited lepton states, such as the excited electron \cite{e*}.
The D0 collaboration searched for associated di-electron production followed by the radiative decay e$\rightarrow e\gamma$
in 1.0 fb$^{-1}$ of data \cite{d0_e*}. The analysis considered single production of an excited electron e* in association with an electron 
via four-fermion contact interactions, with the subsequent electroweak decay of the e* into an electron and a photon.
Figure~\ref{d0_e*} on the left shows the $e\gamma$ invariant mass which shows good agreement with the SM prediction.
Figure~\ref{d0_e*} on the right shows the 95\% C.L. limits for various values of the interaction scale $\Lambda$.
For $\Lambda$ = 1 TeV, masses below 756 GeV are excluded.
\vspace{-0.5cm}
\section{Conclusions}
The CDF and D0 collaborations have an extensive program to look for resonances in the high mass lepton and photon
final states. Analyzing 1.0-2.5 fb$^{-1}$ of data, no significant excess above the standard model expectations is
observed. The derived mass limits on new physics for e.g. new gauge bosons, W' and Z', RS Gravitons, excited electrons
and large extra dimensions are currently the world's most stringent.
\vspace{-0.5cm}
\begin{acknowledgments}
I would like to thank my colleagues in the CDF and D0 collaborations, especially the Exotics and New Phenomena groups for their hard work in 
producing first class results. Thanks to the conference organizers for a superb event.

\end{acknowledgments}


\vspace{-0.5cm}
\begin{thebibliography}{9}   

\bibitem{rs}
L. Randall and R. Sundrum, Phys. Rev. Lett. 83, 4690 (1999); Phys. Rev. Lett. 83, 3370 (1999).
\bibitem{cdf_rs_zprime}
T. Aaltonen $et$ $al.$, arXiv:0810.2059 (2008).
\bibitem{d0_rs}
V.M. Abazov $et$ $al.$, Phys. Rev. Lett. 100, 091802 (2008).
\bibitem{add}
N. Arkani-Hamed, S. Dimopoulos and G. Dvali, Phys. Lett. B {\bf{429}}, 263 (1998).
\bibitem{d0_add}
V.M. Abazov $et$ $al.$, arXiv:0809.2813v1 (2008).
\bibitem{zprime}
J.L. Rosner, Phys. Rev. D 35, 2244 (1987).
\bibitem{d0_wprime}
V.M. Abazov $et$ $al.$, Phys. Rev. Lett. 100, 031804 (2008).
\bibitem{altarelli}
G. Altarelli $et$ $al.$, Z. Phys. C {\bf{45}}, 109 (1989).
\bibitem{e*}
U. Baur, M. Spira and P.M. Zerwas, Phys. Rev. D {\bf{42}}, 815 (1990).
\bibitem{d0_e*}
V.M. Abazov $et$ $al.$, Phys. Rev. D {\bf{77}}, 091102(R) (2008).
\end{thebibliography}
\end{document}